\documentclass[twocolumn,showpacs,preprintnumbers,amsmath,amssymb]{revtex4}

\usepackage{graphicx}
\usepackage{dcolumn}
\usepackage{bm}

\begin{document}

\preprint{PHYSICAL REVIEW B}

\title{Vertical beaming of wavelength-scale photonic crystal resonators}

\author{Se-Heon Kim}
 \email{seheon@kaist.ac.kr}
\author{Sun-Kyung Kim}
\author{Yong-Hee Lee}
\affiliation{%
Department of Physics, Korea Advanced Institute of Science and
Technology, Daejeon 305-701, Korea
}%

\date{\today}

\begin{abstract}
We report that $> 80\%$ of the photons generated inside a photonic
crystal slab resonator can be funneled within a small divergence
angle of $\pm 30^\circ$. The far-field radiation properties of a
photonic crystal slab resonant mode are modified by tuning the
cavity geometry and by placing a reflector below the cavity. The
former method directly shapes the near-field distribution so as to
achieve directional and linearly-polarized far-field patterns. The
latter modification takes advantage of the interference effect
between the original waves and the reflected waves to enhance the
energy-directionality. We find that, regardless of the slab
thickness, the optimum distance between the slab and the reflector
closely equals one wavelength of the resonance under
consideration. We have also discussed an efficient far-field
simulation algorithm based on the finite-difference time-domain
method and the near- to far-field transformation.
\end{abstract}

\maketitle

\section{\label{sec:sec1}Introduction}

Multitudes of optical resonant modes can be created by introducing
a structural defect into a perfect photonic crystal (PhC) having
photonic band gap (PBG). Many researchers have been interested in
designing novel nanocavities based on the PhC in the hope of
controlling the flow of light in unprecedented
ways.\cite{joannopoulos97} Especially two-dimensional (2-D) PhC
slab structures have been widely studied by utilizing
well-established standard fabrication processes.\cite{Johnson99}
In the horizontal directions of such 2-D PhC slab structures,
photons can be spatially localized by 2-D photonic bandgap
effects. Additionally in the vertical direction, photons are
confined rather efficiently through total internal
reflection.\cite{Johnson99,SHKim04} Air-hole triangular lattices
are widely adapted as a basic platform of the 2-D PhC because of
their large photonic bandgap.\cite{Johnson99}

After the first demonstration of the PhC nanocavity laser based on
a single missing air-hole by Caltech group,\cite{Painter99}
various cavity structures have been demonstrated, showing rich
characteristics such as nondegeneracy,\cite{HGPark01}
low-threshold,\cite{HYRyu02} high $Q/V$
value,\cite{Akahane03,HYRyu03} and so on. Recently, 2-D PhC
cavities have drawn much attention as a promising candidate for
cavity quantum electrodynamics (CQED) experiments reporting vacuum
Rabi splitting\cite{Yoshie04} and high-efficiency single photon
sources\cite{Gerard04}. For CQED experiments, various PhC cavities
with fine structural tunings have been proposed to achieve good
coupling between the cavity field and an emitter placed inside the
cavity.\cite{Vuckovic03,Hennessy04} Park \textit{et al.}
demonstrated the first electrically-driven PhC laser based on the
monopole mode that has an electric-field null at the cavity center
where the current-flowing post\cite{HGPark04} is placed. The 2-D
PhC structure can also be used for in-plane integrated optical
circuits where PhC cavities and PhC waveguides are
coupled\cite{Noda02} to supply added functionalities. By using a
PhC-based channel add-drop filter, one can drop (or add) photons
horizontally (or vertically) through the
cavity.\cite{Asano03,Noda00,Takano04} In this way, various
functional PhC devices have been designed.

In this article, we shall discuss useful design rules for the
vertical out-coupling of PhC light emitters. To achieve efficient
out-coupling into a single mode fiber using conventional optics,
the far-field of the PhC resonator should be both well-directed
and linearly polarized. Gaussian-like far-field emission profile
is generally preferable to have good modal overlap with the
fundamental mode of a fiber. Moreover, one needs to find ways to
overcome the strong diffractive tendency of a wavelength-small
cavity. We begin by looking for possible remedies through the
defect engineering approach (tuning the structure near the
defect). Then, we will investigate effects of a bottom reflector
on the far-field characteristics. The surface of high-index ($\sim
3.5$) substrate is regarded as a bottom reflector at first. Note
that such a reflector is practical since it can be naturally
embedded in most of the wafer structures.\cite{HGPark02,HYRyu021}
One can also think of a highly-reflective Bragg mirror made of
alternating layers of high- and low- refractive index dielectric
slabs\cite{Coldren} as an alternative bottom reflector. We will
investigate both the defect engineering approach and the bottom
reflector effects to engineer the far-field radiation
characteristics. A simple and an efficient far-field simulation
method based on the 3-D finite-difference time-domain (FDTD)
method and the fast Fourier transform
(FFT)\cite{Taflove,Vuckovic02} will also be discussed.

This article is organized as follows. In Sec.~\ref{sec:sec2},
various characteristics of 2-D PhC slab resonant modes are
reviewed. In Sec.~\ref{sec:sec3}, we will describe the far-field
simulation method and show simulation results. In
Sec.~\ref{sec:sec4}, we will discuss how the defect engineering
method can be applied to the PhC resonant mode to realize
directional emission. Then, effects of the presence of a bottom
reflector will be discussed. Finally, we discuss how the two
proposed schemes can be combined to give both directional and
linearly-polarized emission.

\section{\label{sec:sec2}Resonant modes in a modified single defect cavity}

\begin{figure}
\includegraphics{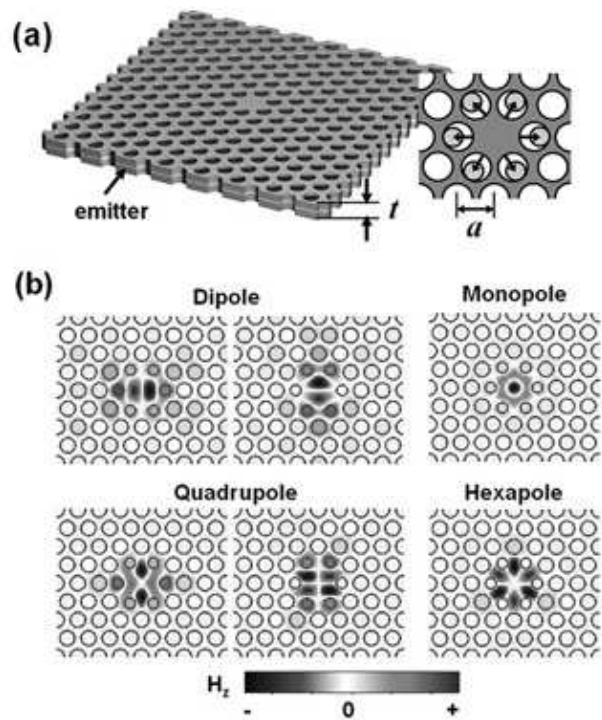}
\caption{\label{fig:fig1} (a) Structure of a modified
single-defect cavity. Here, the six nearest neighbor holes are
pushed away from the cavity center after reducing their radii from
$r=0.35 a$ to $r_m= 0.25 a$. (b) Various resonant modes in a
modified single-defect cavity. Dipoles and quadrupoles are doubly
degenerate while a monopole and a hexapole are nondegenerate. We
plot $H_z$ components of the each resonant mode, where the
electric-fields are polarized in the $x$-$y$ plane (TE-like
modes).}
\end{figure}

In this section, we review some important characteristics of a few
possible resonant modes in a modified single defect PhC
cavity.\cite{HGPark02} The 3-D FDTD method was employed to analyze
optical characteristics such as the mode profile, the $Q$ factor,
the resonant frequency, and so on. A typical structure of the
modified single defect cavity and various resonant modes are shown
in Fig.~\ref{fig:fig1}(a). A free-standing PhC slab structure is
assumed as a platform for a triangular PhC lattice. The refractive
index of the slab is chosen to be 3.4, which corresponds to that
of GaAs around 1 $\rm{\mu}m$. For real applications, one can
insert active semiconductor layers, such as the InAs quantum dots,
in the middle of the slab. The air-hole radius and the slab
thickness are chosen to be 0.35$a$ and 0.50$a$, respectively,
where $a$ represents the lattice constant of the triangular PhC.
The choice of the slab thickness is important to support only
fundamental modes of the slab (TE-like modes).\cite{HYRyu00}
Without modification on the nearest air-holes around the defect,
only the doubly-degenerate dipole mode can be
excited.\cite{Painter99,Painter991} However, by reducing the
radius of the six nearest air-holes as indicated in
Fig.~\ref{fig:fig1}(a), other resonant modes within the PBG can be
excited. This type of modification was first proposed by Park
\textit{et al.} so as to find a nondegenerate monopole mode used
for a photo-pumped low-threshold PhC laser.\cite{HGPark01}
Especially, as indicated by Ryu \textit{et al.}, the hexapole mode
can have a very high $Q/V$ value due to the whispering gallery
nature of the symmetric field distribution.\cite{HYRyu03} For
those modes having an electric-field intensity node at the cavity
center, intrinsic properties of those resonant modes is not
changed appreciably even after the introduction of a small post at
the cavity center. This was a crucial design rule applied for the
electrically-pumped PhC laser.\cite{HGPark04} One disadvantage of
these resonant modes is that the vertical emission is inherently
prohibited due to the destructive interference of the
anti-symmetric field patterns.

Resonant frequencies and $Q$ factors of possible resonant modes in
the modified single defect cavity are summarized in
Table~\ref{tab:table1}. Generally, the total radiated power
($1/Q_{tot}$) can be decomposed into a vertical contribution
($1/Q_{vert}$) and an in-plane contribution
($1/Q_{horz}$).\cite{Painter991} One usually refers to
$1/Q_{vert}$ as the inherent optical loss of the resonant mode,
because the in-plane loss ($1/Q_{horz}$) can be arbitrary reduced
by increasing the number of PhC layer surrounding the cavity. In
fact, the $Q$ factors listed in Table~\ref{tab:table1} are
$Q_{tot}$, which have been obtained by using a sufficiently large
horizontal computational domain ($16 a \times 16 a$). The optical
loss of the resonant mode is closely related to the in-plane field
distributions. For graphical illustration of the vertical loss,
the momentum space intensity distribution\cite{Srinivasan02}
obtained by Fourier transforming the in-plane field
($|FT(E_x)|^2+|FT(E_y)|^2$) is calculated. From the in-plane
momentum conservation rule, only plane-wave components inside a
light-cone [$k_x^2+k_y^2 = (\omega/c)^2$] can couple with
propagation modes.

\begin{table}
\caption{\label{tab:table1}Normalized frequencies ($\omega_n
=a/\lambda$), $Q$ factors, and theta-polarized percentages of the
resonant modes shown in Fig.~\ref{fig:fig1}(b).}
\begin{ruledtabular}
\begin{tabular}{cccc}
~& $\omega_n$& $Q$ & $P_{\theta}/P_{total} (\%)$\\
\hline
Dipole & 0.2886 & 14 900 & 46\\
Quadrupole & 0.3179 & 45 000 & 75.9\\
Monopole & 0.3442 & 11 000 & 1.2\\
Hexapole & 0.3149 & 168 000 & 74.5
\end{tabular}
\end{ruledtabular}
\end{table}

\begin{figure}
\includegraphics{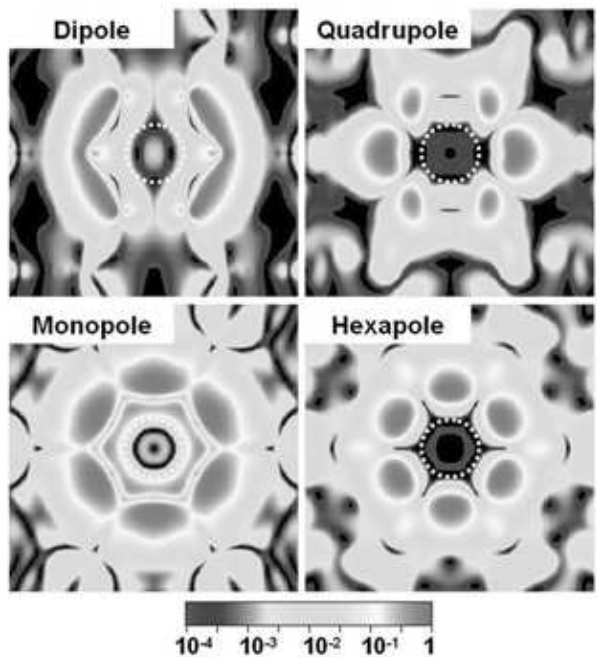}
\caption{\label{fig:fig2} Momentum space intensity distribution
($|FT(E_x)|^2+|FT(E_y)|^2$) of the resonant modes in a modified
single-defect cavity. The dotted white circle in each graph
represents a light-line defined by $k_x^2+k_y^2 = (\omega/c)^2$.
The major fraction of the plane-wave components is outside the
light-line, which implies the efficient index confinement
mechanism.}
\end{figure}

As can be seen from the momentum space intensity distributions of
the resonant modes in Fig.~\ref{fig:fig2}, the major fraction of
the resonant photons live outside the light-cone, which implies an
efficient index confinement mechanism. The hexapole mode has
negligible plane-wave components inside the light-cone, hence a
high $Q$ factor of $\sim$ 168 000. Among four possible resonant
modes, the monopole mode and the dipole mode have relatively low
$Q$ factors ($\sim$ 10 000). Due to the presence of dc components
($k_x$ = $k_y$ = 0), the dipole mode shows relatively large
radiation losses. Although the controlled vertical emission is
essential for good out-coupling, one should not give up the
quality factor too much. In fact, several groups have reported
high $Q/V$ PhC cavities with directional
emission.\cite{Akahane03,Vuckovic03,Noda02} Later, we will
investigate how one can perform additional fine tunings on the
modified single defect cavity to obtain both directionality and
high $Q/V$.
\begin{figure*}
\includegraphics{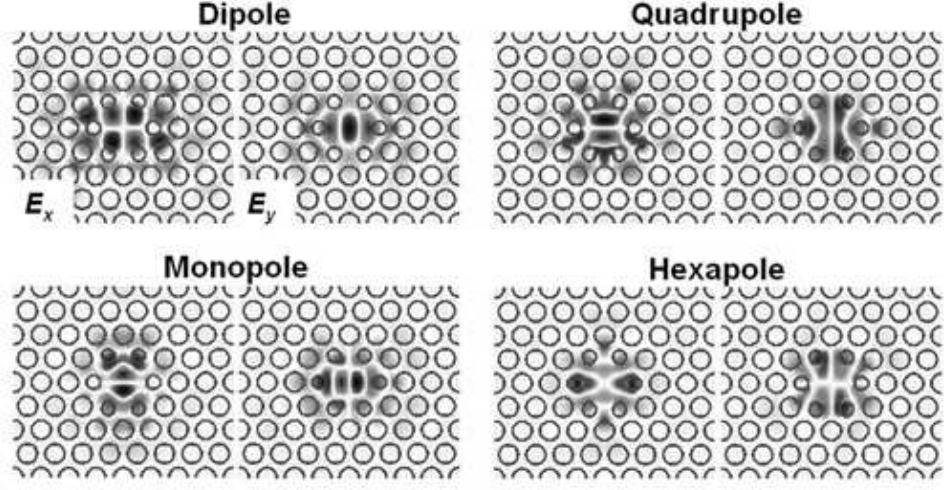}
\caption{\label{fig:fig3} $E_x$ and $E_y$ field distributions of
the resonant modes in a modified single-defect cavity. For the
doubly degenerate modes such as the dipole and the quadrupole, we
plot one of the degenerate pair (the left ones among degenerate
pairs shown in Fig.~\ref{fig:fig1}).}
\end{figure*}

Finally, we would like to mention how the structural symmetry of
the PhC cavity can affect far-field characteristics. The modified
single defect PhC cavity depicted in Fig.~\ref{fig:fig1} retains
complete six-fold symmetry. The group theory tells us one can
classify all the resonant modes into nondegenerate modes and
doubly degenerate modes depending on the rotational symmetry; the
monopole mode and the hexapole mode are nondegenerate and the
dipole mode and the quadrupole mode are doubly
degenerate.\cite{SHKim03,Sakoda} Fig.~\ref{fig:fig3} shows $E_x$
and $E_y$ field distributions of the resonant mode. First, let us
look at the case of one of the degenerate dipole modes. In $E_x$
field distribution, due to the odd reflection symmetries by both
$x=0$ and $y=0$, the $E_x$ vertical emission is prohibited.
However, in $E_y$ distribution, since all the reflection
symmetries are even, there remains a dc component which
contributes $y$-polarized vertical emission in the far-field. In
the cases of the monopole mode and the quadrupole mode, at least
one of the reflection symmetries are odd, thus the vertical
emission is always prohibited. In the case of the hexapole mode,
$E_x$ field distribution shows even reflection symmetries by both
axes ($x=0$ and $y=0$). Considering its perfect six-fold symmetry,
$E_x$ field components must be completely balanced to give a zero
dc component. Thus, $E_x$ field distribution reveals that such
delicate balance can be easily broken by introducing a small
structural perturbation, resulting in nontrivial vertical
emission. We will discuss this type of defect engineering in
Sec.~\ref{sec:sec4}.

\section{\label{sec:sec3}Far-field simulation}

\subsection{\label{sec:sec3p1}Far-field simulation method}

To obtain the far-field radiation pattern, we should deal with the
radiation vectors that are sufficiently far ($r\gg\lambda$) from a
light emitter.\cite{Jackson} Direct application of the FDTD method
would be difficult because of the limited computer memory size and
time. Here, we shall explain how one can efficiently obtain the
far-field radiation by combining the 3-D FDTD method and the near-
to far-field transformation formulae. This method has been well
known in the field of electronics since 1980s dealing with antenna
radiation problems. V\v{u}kovi\'{c} \textit{et al.} applied the
FFT-based far-field simulation algorithm to their $Q$ factor
optimization of the PhC cavity mode.\cite{Vuckovic02}

We explain the basic concept of the efficient far-field
computation in Fig. 4. Here, we divide the whole calculation
domain by a horizontal plane located just above the PC slab. The
upper domain is bounded by an infinite hemisphere. All field
components (${\bf E}$ and ${\bf H}$ ) are assumed to fall off as
$1/r$, typical of radiation fields. It is also assumed that
in-plane field components ($E_x$, $E_y$, $H_x$, and $H_y$)
detected at the horizontal plane decay exponentially as the
horizontal distance $\rho$. According to the surface equivalence
theorem, for the calculation of the fields inside an imaginary
closed space $\Omega$, the equivalent electric (${\bf J}_s$) and
magnetic (${\bf M}_s$) currents on the surrounding surface can
substitute all the information on the fields out of the $\Omega$,
where ${\bf J}_s$ and ${\bf M}_s$ are calculated by using the
following formulae.

\begin{eqnarray}
{\bf J}_s = \hat{n} \times {\bf H} = -\hat{x} H_y + \hat{y} H_x
\label{eq:3p1}\\
{\bf M}_s = -\hat{n} \times {\bf E} = \hat{x} E_y - \hat{y} E_x
\label{eq:3p2}
\end{eqnarray}

Here, $\hat{n}$ is a unit normal vector on the surface. Neglecting
the fields on the hemisphere, only the in-plane field data
detected at the horizontal plane will be used for the far-field
computation. Then, these equivalent currents are used to obtain
the following retarded vector potentials.
\begin{figure}
\includegraphics{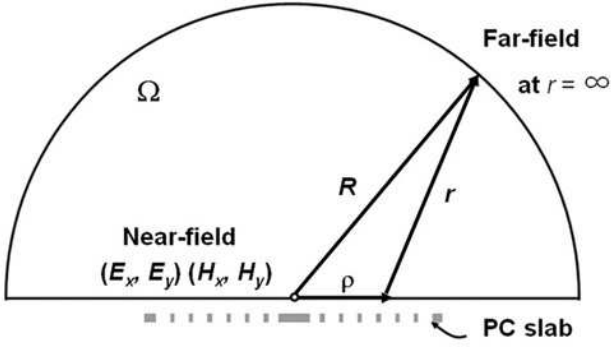}
\caption{\label{fig:fig4} Geometry used for the far-field
computation. The whole calculation domain is divided by the
horizontal plane just above the photonic crystal slab. According
to the surface equivalence theorem, the near-field components
($E_x$, $E_y$, $H_x$, and $H_y$) at the horizontal plane give
sufficient information on the far-field pattern in the upper
domain ($\Omega$).}
\end{figure}

\begin{eqnarray}
{\bf A} = \mu_o \int_s \frac{{\bf J}_s e^{-ikr}}{4 \pi r} dS
\label{eq:3p3} \\
{\bf F} = \epsilon_o \int_s \frac{{\bf M}_s e^{-ikr}}{4 \pi r} dS
\label{eq:3p4}
\end{eqnarray}

In the far-field regime, ($r \approx R - \rho \cos \Psi$), the
above formulae can be simplified as

\begin{eqnarray}
{\bf A} = \mu_o \frac{e^{-ikR}}{4 \pi R} \int_s {\bf J}_s e^{ik
\rho \cos \Psi} dS \equiv \mu_o \frac{e^{-ikR}}{4 \pi R} {\bf N}
\label{eq:3p5}\\
{\bf F} = \epsilon_o \frac{e^{-ikR}}{4 \pi R} \int_s {\bf M}_s
e^{ik \rho \cos \Psi} dS \equiv \epsilon_o \frac{e^{-ikR}}{4 \pi
R} {\bf L} \label{eq:3p6}
\end{eqnarray}

Using a relation $k \rho \cos \Psi = k_x x + k_y y$, the radiation
vectors ${\bf N}$ and ${\bf L}$ are simply related by 2-D FTs.

\begin{eqnarray}
{\bf N} \equiv \int_s {\bf J}_s e^{ik \rho \cos \Psi} dS = FT(
{\bf J}_s ) \label{eq:3p7} \\
{\bf L} \equiv \int_s {\bf M}_s e^{ik \rho \cos \Psi} dS = FT(
{\bf M}_s ) \label{eq:3p8}
\end{eqnarray}

$x$ and $y$ components of the radiation vectors can be represented
as

\begin{eqnarray}
N_x (k_x, k_y) = -FT \{ H_y (x,y) \} \label{eq:3p9} \\
N_y (k_x, k_y) = +FT \{ H_x (x,y) \} \label{eq:3p10} \\
L_x (k_x, k_y) = +FT \{ E_y (x,y) \} \label{eq:3p11} \\
L_y (k_x, k_y) = -FT \{ E_x (x,y) \} \label{eq:3p12}
\end{eqnarray}

Then, far-field radiation patterns are obtained by calculating
time-averaged Poynting energy per unit solid angle.\cite{Jackson}

\begin{equation}
\frac{dP}{d\Omega} = R^2 \langle {\bf E} \times {\bf H} \rangle
\cdot \hat{n} = \frac{k^2 \eta}{32 \pi^2} \left( \left|
N_{\theta}+\frac{ L_{\phi}}{\eta} \right|^2 + \left|
N_{\phi}-\frac{ L_{\theta}}{\eta} \right|^2 \right)
\label{eq:3p13}
\end{equation}
where $\eta$ is the impedance of free space.

One can easily implement this algorithm into the FDTD code. During
the FDTD time stepping, the in-plane field data (near-field data)
that contain complete information on the far-field are obtained.
Once the FDTD computation is completed, the radiation vectors
(${\bf N}$ and ${\bf L}$) are calculated by the FTs. Then, these
radiation vectors are used to obtain far-field patterns according
to Eq.~(\ref{eq:3p13}). One can use the 2-D FFT algorithm for
efficient FT computations [see
Eqs.~(\ref{eq:3p9})-(\ref{eq:3p12})]. In our FDTD simulation, the
size of the computational grid was $a/20$ ($a$ is the lattice
constant), and the size of the horizontal calculation domain were
$16 a \times 16 a$. Thus, each in-plane field data will contain
approximately $320 \times 320$ data points. In the FFT algorithm,
$N \times N$ input data ${\cal I}(x,y)$ are transformed to the
output matrix data ${\cal O}(f_x,f_y)$ with the same $N \times N$
size, where $f$ is $k/2\pi$.\cite{Press} Assume that $N$ unit
cells in the ${\cal I}(x,y)$ data correspond to a spatial length
of $L$ wavelengths ($\lambda$). The frequency step $\triangle f$
is $1/(L \lambda) \equiv f_0 /L$ and the maximum frequency is thus
$Nf_0/L$. Thus, the value $f=f_0$ of the light-cone corresponds to
the Lth point in the frequency space. Note that only the
wavevector components lying inside the light-cone whose radius is
$L$ cells can contribute to the outside radiation. To increase the
resolution of the far-field simulation, one should increase the
horizontal size of the FDTD domain. Fortunately, there is an easy
way to do so without performing an actual calculation in the
larger FDTD domain. We can enlarge the input matrix data by
filling null (0) values, for example, into $2048 \times 2048$ size
matrix. Such arbitrary expansion is legitimate since most of the
near-field energy of a PhC cavity mode is strongly confined around
the central defect region. When the normalized frequency of a PhC
cavity mode is $a/\lambda \sim 0.35$ (typical of PhC defect
modes), $L$ = 2048 cells = 102.4$a$ $\sim 36 $($\lambda$). Thus,
the resultant far-field data resolution is $\triangle f = f_0/L
\sim f_0/36$.

Finally, we would like to mention one more before presenting
calculated far-field patterns. All ${\bf E}$ and ${\bf H}$ data
described in the above formulae are phasor quantities. Thus, when
we detect the in-plane field data, we have to extract both
amplitude and phase information, because, in general, phases will
change differently at each position. The situation containing the
bottom reflector is the case, where the reflected waves can no
longer be described as a simple standing wave of ${\bf E}({\bf
r},t)=|{\bf E}({\bf r})|e^{i\omega t}$. Assuming a single mode
$\omega$, all field vectors in the simulation should have the
following form.

\begin{equation}
{\bf E}({\bf r},t)= \tilde{{\bf E}}({\bf r}) \exp(i\omega t) =
|{\bf E}({\bf r})|\exp(i \varphi ({\bf r})) \exp(i\omega t)
\label{3p14}
\end{equation}

An efficient method to obtain the exact phasor quantities
$\tilde{{\bf E}}({\bf r})$ is to incorporate recursive discrete
FTs (DFTs) concurrently with the FDTD time stepping.

\begin{eqnarray}
{\rm Ex_{real}}|_{i,j}^n = {\rm Ex_{real}}|_{i,j}^{n-1} + E_x
|^n_{i,j} \cdot \cos( \omega
\cdot n \Delta t ) \label{3p15}\\
{\rm Ex_{imag}}|_{i,j}^n = {\rm Ex_{imag}}|_{i,j}^{n-1} + E_x
|^n_{i,j} \cdot \sin( \omega \cdot n \Delta t ) \label{3p16}
\end{eqnarray}
Here, to obtain real (${\rm Ex_{real}}$) and imaginary (${\rm
Ex_{imag}}$) components of $E_x$ field, simple recursive
summations are performed on each grid point ($i,j$) at the FDTD
time $n$. Note that the above formulae correspond to the following
Fourier transformation.

\begin{equation}
\tilde{E_x}({\bf r},\omega) = \int E_x ({\bf r},t) e^{i \omega t}
dt \label{3p16a}
\end{equation}

In general, this summation should continue until the field values
`ring down' to zero. However, in the case of high $Q$ factor
resonant mode, by carefully exciting a dipole source (source
position and frequency), small FDTD time steps (usually 20 000
$\sim$ 30 000 FDTD time steps, or equivalently $\sim$ 200 optical
periods) are sufficient to obtain nearly exact phasor quantities.
In the next section, we will show calculated far-field patterns.

\subsection{\label{sec:sec3p2}Far-field simulation results}
\begin{figure}
\includegraphics{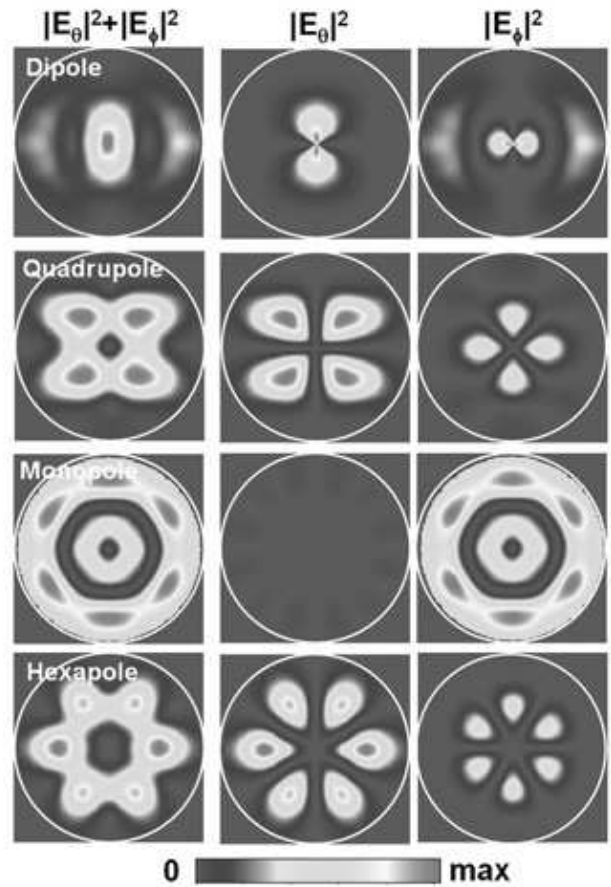}
\caption{\label{fig:fig5} Calculated far-field patterns of the
resonant modes depicted in Fig. 3. All the far-field data ($x,y$)
are represented by using a simple mapping defined by $x=\theta
\cos \phi$ and $y=\theta \sin \phi$ (The radius of the plot
corresponds to $\theta$.). Along with total intensity ($|{\bf
E}|^2$) patterns, we also display theta-polarized intensity
($|E_{\theta}|^2$) patterns and phi-polarized intensity
($|E_{\phi}|^2$) patterns. Intensities of $|E_{\theta}|^2$ and
$|E_{\phi}|^2$ are normalized by the maximum value of both data.}
\end{figure}

To calculate the far-field pattern, we have used the 3-D FDTD
method and the near- to far-field transformation algorithm which
has been outlined in the previous subsection. The horizontal
calculation domain of $16 a \times 16 a$ and the spatial grid
resolution of $a/20$ were used. Calculated far-field patterns are
shown in Fig.~\ref{fig:fig5}. All the far-field data ($x,y$) shown
in this article are represented by using a simple mapping defined
by $x=\theta \cos \phi$ and $y= \theta \sin \phi$. Depending on
the rotational symmetry that the each resonant mode has, the
far-field pattern also shows several regularly spaced lobes, which
in turn explain the labeling of these modes; Number of lobes in
the far-field pattern are four (six) in the case of the quadrupole
mode (the hexapole mode). Ignoring weak intensity modulation laid
upon dominating circular emission pattern, the monopole mode shows
two concentric ring-shape patterns, which show indeed a
monopole-like oscillation. In fact, emission patterns of the
hexapole mode and the monopole mode does not bear perfect six-fold
symmetry, e.g. two horizontal lobes look bigger than the others in
the case of the hexapole mode. These discrepancies are due to
rectangular grids of the FDTD computation cell which are not well
matched with the six-fold symmetric case. (Later, we will explain
more in detail.) Along with total intensity ($|{\bf E}|^2$)
patterns, we also display theta-polarized intensity
($|E_{\theta}|^2$) patterns and phi-polarized intensity
($|E_{\phi}|^2$) patterns. Note that such polarization-resolved
far-field data can be easily obtained in the real measurement
setup (a solid angle scanner) by placing a polarizer in front of a
photo-detector which scans the whole hemispherical
surface.\cite{DJShin02}

First, let us look at the results of the quadrupole mode.
Comparing with the total intensity power, it can be deduced that
most of the emitted power is theta-polarized. Also in the case of
the hexapole mode, theta-polarized emission seems to dominate
phi-polarized one. Table~\ref{tab:table1} summarize percentages of
the theta-polarized power and the phi-polarized power. As
expected, the quadrupole mode and the hexapole mode are almost
theta-polarized ($>70\%$). However, the monopole mode shows
opposite result; phi-polarized emission dominates by $\sim 99\%$.
As indicated by Povinelli \textit{et al.}, one may create an
artificial magnetic emitter by using the monopole-like mode which
shows the same emission characteristics reminiscent of those of
the magnetic multipole.\cite{Povinelli03} Considering that
electric (magnetic) dipole radiation is theta-polarized
(phi-polarized),\cite{Jackson} such polarization characteristics
seem to be closely related with composition of the electric
multipoles and the magnetic multipoles. Thus, the so called
multipole decomposition could provide other perspectives for
various radiation properties of PhC nanocavities along with the
plane-wave decomposition (FT based analyses).

\begin{figure}
\includegraphics{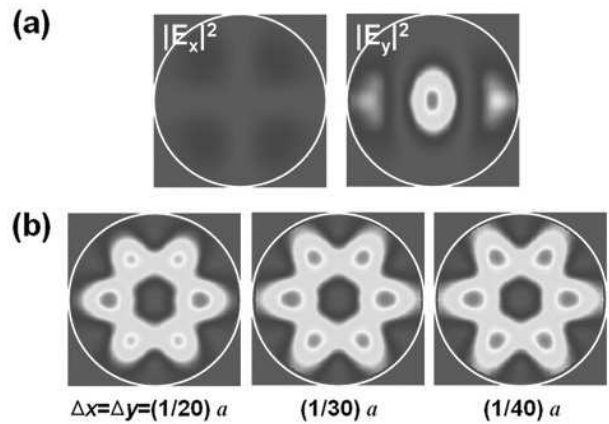}
\caption{\label{fig:fig6} (a) $x$-polarized intensity ($|E_x|^2$)
pattern and the $y$-polarized intensity ($|E_y|^2$) pattern of the
dipole mode. Intensities of $|E_x|^2$ and $|E_y|^2$ are normalized
by the maximum value of both data. (b) Calculated far-field
patterns of the hexapole mode for the finer grid resolution
($\Delta x$, $\Delta y$). The vertical grid resolution $\Delta z$
was fixed to be $a/20$ throughout the calculation.}
\end{figure}

From polarization-resolved far-field patterns, it looks as if the
dipole mode has a certain singularity near the north pole
($\theta=0$). This feature comes from the strongly $y$-polarized
dipole mode. Upon inspection of the region around the vertical,
one can observe that directions of the electric field vectors
indeed predominantly lie in the $y$ direction.
Fig.~\ref{fig:fig6}(a) shows the $x$-polarized emission pattern
($|E_x|^2$) and the $y$-polarized emission pattern ($|E_y|^2$).
The result shows the domination of the $y$-polarized emission over
the $x$-polarized one.\cite{Painter02} Thus, except the inherent
degeneracy, the dipole mode itself can be a good candidate for the
vertical emitter.

Finally, we would like to mention the origin of the unevenness for
the six lobes in the far-field patterns of the hexapole mode and
the monopole mode. These symmetry-breaking comes from the
rectangular grid, used in the FDTD computation, which does not fit
in the hexagonal symmetry. To check this point, we re-calculated
the far-field pattern of the hexapole mode with the grid of the
finer grid resolution. In this computation, $\Delta z$ was fixed
and only the horizontal grids ($\Delta x, \Delta y$) were changed.
Fig.~\ref{fig:fig6}(b) shows that, by increasing the horizontal
grid resolution, the far-field pattern restores the inherent
six-fold symmetry. This comparison confirms that such unbalanced
lobes are attributable to the rectangular FDTD grid. Note that
even in case of $\Delta x = \Delta y = a/20$, all important
features of the radiation can still be obtained except the slight
unevenness. Thus, all the calculations in Sec.~\ref{sec:sec4} were
performed with the grid resolution of $a/20$.

\section{\label{sec:sec4}Engineering radiation properties}

\subsection{\label{sec:sec4p1}Defect engineering}

\begin{figure}
\includegraphics{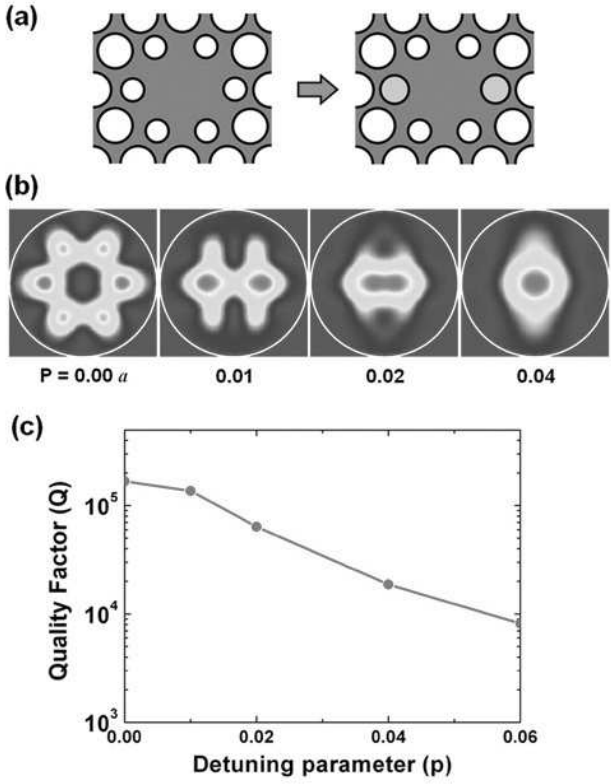}
\caption{\label{fig:fig7} Radiation characteristics of the
deformed hexapole mode. (a) Schematic of the deformed structure,
where two horizontal air-holes facing with each other are enlarged
by $p(a)$. Except for this enlargement, all the structural
parameters are kept the same as the previous structure shown in
Fig.~\ref{fig:fig1}(a). (b) Calculated far-field patterns for
various values of the air-hole size increment $p$. When $p = 0.04
a$, the far-field pattern shows good directionality. (c) Quality
factor of the deformed hexapole mode as a function of the
deformation parameter $p$.}
\end{figure}

As discussed, the in-plane field distribution determines the
far-field radiation pattern. In this subsection, we will
investigate how the defect engineering method can be applied to
manipulate the radiation profile. The hexapole mode depicted in
Sec.~\ref{sec:sec2} shows null vertical emission due to the
symmetric field distribution which has a zero dc component.
However, as discussed in the last part of Sec.~\ref{sec:sec2}, it
seems that the field balance can be easily broken by introducing
small structural perturbation. As shown in Fig.~\ref{fig:fig7}(a),
the structural tuning that we have applied is to enlarge two
horizontal air-holes facing with each other.\cite{SKKim} From the
hexapole mode field distribution shown in Fig. 3, it can be
expected that such perturbation will break the delicate balance of
the $E_x$ distribution. Note that noticeable amount of the dc
component appears in the $E_x$.\cite{SHKwon}

In Fig.~\ref{fig:fig7}(b), the far-field patterns are plotted as
we increase the air-hole radius. Here the same structural
parameters used in Fig.~\ref{fig:fig1} are assumed. One can
clearly observe how the hexagonally-distributed pattern is
transformed into the directional beaming. The radiation pattern
becomes more directional with the air-hole radius. When the
air-hole radius increment is $0.04 a$, the vertical emission looks
most favorable. In this case, $\sim 65\%$ of the total radiated
energy falls within the collection angle of $\pm 30^{\circ}$. In
other words, $\sim 32.5\%$ of the total emitted photons can be
collected from the top. The downward photons could be re-directed
by introducing a reflector underneath. In the next section, we
shall discuss the efficient unidirectional beaming strategy based
on the defect tuning and the bottom reflector. The $Q$ factors of
the proposed hexapole mode are presented in
Fig.~\ref{fig:fig7}(c). The exponential decrease of the $Q$ factor
is observed as a function of the detuning parameter ($p$). Note
that when the $p = 0.04 a$, the $Q$ factor is still not too bad,
$\sim$ 20 000.

\begin{figure}
\includegraphics{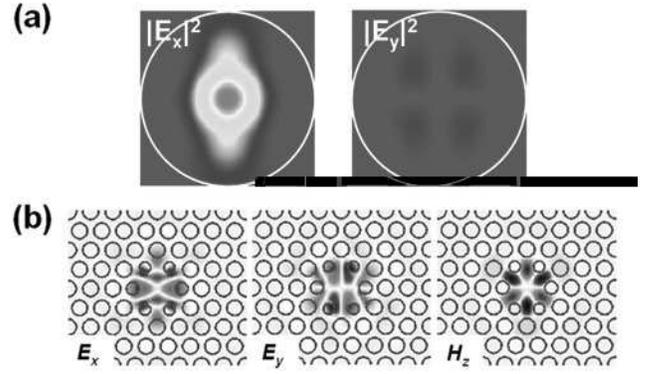}
\caption{\label{fig:fig8} (a) $x$-polarized intensity ($|E_x|^2$)
pattern and the y-polarized intensity ($|E_y|^2$) pattern of the
deformed hexapole mode. Intensities of $|E_x|^2$ and $|E_y|^2$ are
normalized by the maximum value of both data. (b) $E_x$, $E_y$,
and $H_z$ field components of the deformed hexapole mode when
$p=0.04 a$. Note that, although the resulting near-field
distribution is almost identical to that of the ideal hexapole
mode, the far-field pattern changes drastically.}
\end{figure}

Since the above modification of two air holes breaks mostly the
balance of the $E_x$ field and not that of the $E_y$ field, the
directional beaming obtained here will be almost $x$-polarized.
Fig.~\ref{fig:fig8}(a) shows the polarization-resolved far-field
patterns, where such directional beaming is clearly $x$-polarized.
With the same numerical aperture of the lens (0.5) used
previously, the polarization extinction ratio is estimated to be
$\sim 70:1$. Therefore, we claim that both the directionality and
the linear polarization can be obtained from the hexapole mode by
introducing small structural perturbation. However, by the same
token, this result implies that the hexapole mode in the modified
single defect cavity is not robust against the structural
perturbation, because the electric fields are concentrated in the
proximity of the nearest air-holes. This fact also explains one of
the reasons why the observed $Q$ factors of the hexapole mode are
much smaller than the values obtainable in the FDTD computation
.\cite{HYRyu03,GHKim} Recently the other class of hexapole modes
that are robust against the perturbation was also found, where the
electric field energy is concentrated in the dielectric region
between the nearest air-holes.\cite{SHKwon,SKKim05}

Finally, it should be noted that the near-field distribution of
the deformed hexapole mode is almost identical to that of the
ideal hexapole mode and retains the whispering-gallery-like
nature. Remember that the $E_x$, $E_y$, and $H_z$ field components
shown in Fig.~\ref{fig:fig8}(b) is the results obtained after only
very slight modification of the holes ($p=0.04 a$). It is
interesting that the resulting far-field pattern can be changed
drastically by such minute perturbation. Thus, as indicated by
Shin \textit{et al.} a thorough understanding of wavelength-size
small cavities requires investigations in the both regimes; the
near-field and the far-field.\cite{DJShin02}

\subsection{\label{sec:sec4p2}Effects of the bottom reflector}

Purcell claimed that the spontaneous emission lifetime is a
constant that can be altered depending on the optical mode density
and the electric-field intensity near the emitter
.\cite{Purcell46} Borrowing this concept, it can be said that
inherent radiation properties of the PhC slab resonant mode should
be modified by placing a reflector near the cavity.\cite{Bjork91}
In the early days, researchers used PhC resonant cavities as a
vehicle to study the alteration of spontaneous emission
.\cite{JKHwang99,RKLee00,HYRyu031} For example inside the periodic
structure, the spontaneous emission can be suppressed for a
certain frequency range (PBG) or enhanced for a particular
frequency. As will be shown below, one can further modify the
resulting radiation properties from the PhC cavity by using a
flat-surface mirror.\cite{Hinds}

\begin{figure}
\includegraphics{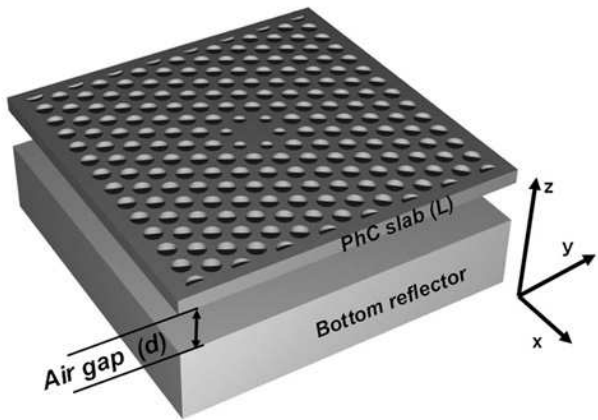}
\caption{\label{fig:fig9} Schematic of a novel photonic crystal
unidirectional emitter which consists of a photonic crystal slab
and a bottom reflector. The resonant modes are confined around the
defect region of the slab.}
\end{figure}

\begin{figure}
\includegraphics{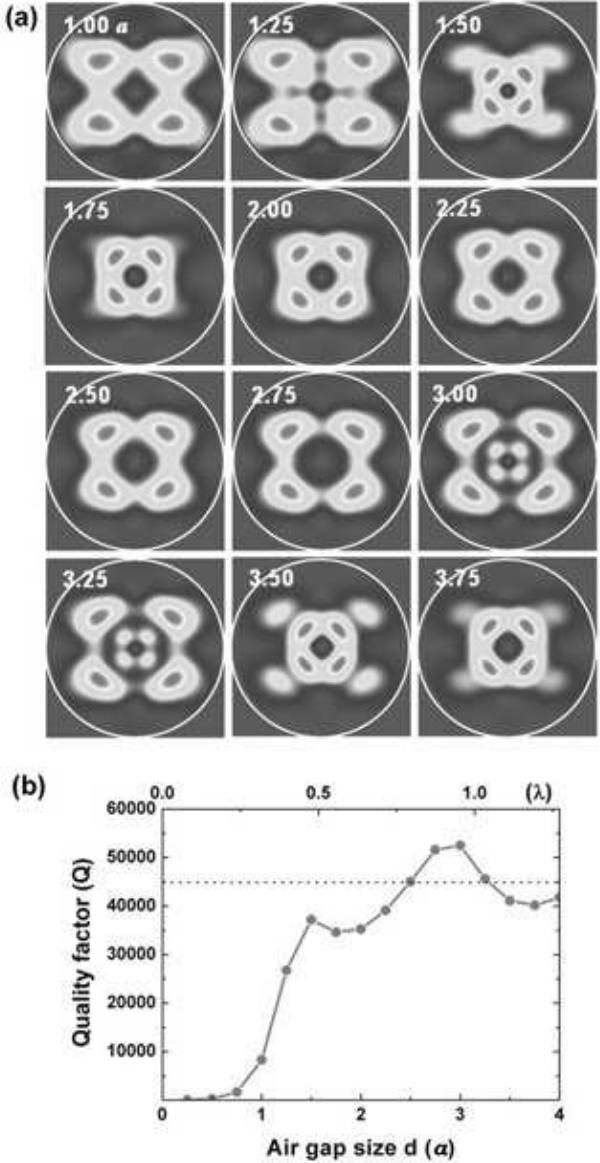}
\caption{\label{fig:fig10} Radiation characteristics of the
quadrupole mode in the presence of a simple dielectric substrate
with a refractive index of 3.4. (a) Calculated far-field patterns
for various values of the air-gap thickness. Here, each intensity
value has been normalized to the maximum value of its own data.
(b) Calculated quality factors as a function of the air-gap size.
A horizontal dotted line represents the original quality factor in
the absence of the reflector.}

\end{figure}

The PhC slab structure to be investigated here is shown in
Fig.~\ref{fig:fig9}. The proposed structure is very simple; it
consists of a PhC slab and a bottom reflector. As a reflector, a
simple dielectric substrate or a Bragg mirror is to be used. Such
a structure is realistic since the thickness of each layer can be
controlled during the wafer growth. We note that similar planar
systems were studied by Schubert\cite{Schubert92} and
Benisty\cite{Benisty98} in an attempt to alleviate the poor light
extraction efficiency of microcavity LEDs. To investigate effects
of the presence of the bottom reflector, we have calculated the
change of $Q$ factors and far-field radiation patterns as we vary
the air-gap distance between the PhC slab and the bottom reflector
(see Fig.~\ref{fig:fig10}). Here, the $Q$ factor represents the
lifetime of the resonant mode ($\tau =Q/\omega$). As a PhC
resonant mode, the quadrupole mode shown in Fig.~\ref{fig:fig5} is
used. As a bottom reflector, we assumed a simple dielectric
substrate whose refractive index is 3.4.

Compared with the original far-field pattern shown previously in
Fig.~\ref{fig:fig5}, the radiation profile shown in
Fig.~\ref{fig:fig10} changes noticeably with the air-gap size. The
far-field patterns show the signature of concentric ring-shape
modulation laid upon the original far-field pattern, which is
reminiscent of the Fabry-P\'{e}rot fringes \cite{Born} in the
Fabry-P\'{e}rot etalon. Although the reflectivity of the bottom
GaAs (or InP) substrate is only $\sim 30\%$, the interference
between the original wave and the reflected wave is easily
noticeable. Upon inspection of the far-field patterns in
Fig.~\ref{fig:fig10}(a), one can find that the similar far-field
patterns show up again after the air-gap increment of $\sim 1.75
a$ (for example, compare $1.75 a$ and $3.50 a$). This increment
corresponds to around a half-wavelength ($\sim 1.75 a \times
0.3179 \lambda / a \sim 0.55 \lambda$). It should be noted that
the directional beam cannot be achieved from the quadrupole mode.
To achieve unidirectional emission, we need to investigate the
inherently directional resonant mode such as the deformed hexapole
mode.
\begin{figure}
\includegraphics{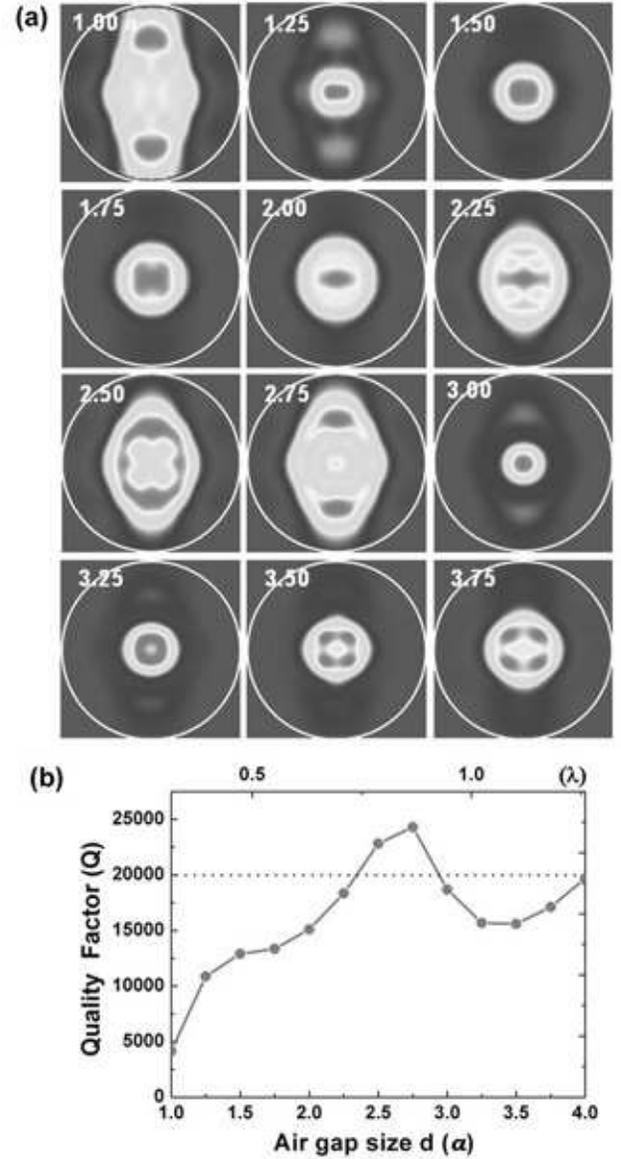}
\caption{\label{fig:fig11} Radiation characteristics of the
deformed hexapole mode in the presence of a simple dielectric
substrate with a refractive index of 3.4. (a) Calculated far-field
patterns for various values of the air-gap thickness. Here, each
intensity value has been normalized to the maximum value of its
own data. (b) Calculated quality factors as a function of the
air-gap size. A horizontal dotted line represents the original
quality factor in the absence of the reflector.}
\end{figure}

Similar systematic dependence on the air-gap size is also observed
in the $Q$ factor of the resonator, at a period of approximately
$\lambda/2$. Note that the $Q$ factor decreases drastically when
the air-gap size becomes less than $0.25 \lambda$. There are two
distinct reasons for this phenomenon; one is the TE-TM coupling
loss through the PhC slab \cite{Tanaka03} and the other is the
diffraction loss in the downward direction. The TE-TM coupling
loss always occurs when the vertical symmetry of the PhC slab
structure is broken. The reason for the increase of the
diffraction loss to the downward direction is that the light-cone
of the bottom substrate is effectively enlarged by a factor of $n$
(the refractive index of the substrate). When the bottom substrate
is too close from the slab, the tail of the evanescent field
begins to feel the substrate. Such an expanded light-cone will
allow the more plane-waves to be coupled to substrate-propagating
modes. Therefore, in order to prevent this loss, the air-gap size
larger than $\lambda/2$ is generally recommended. It is
interesting to find that the $Q$ factor can be enhanced even in
the presence of the bottom substrate. In fact, when the air-gap
size is larger than $\lambda/2$, the $Q$ factor oscillates around
the original value. Note that the interference phenomena observed
here are similar to the cases of the dipole antenna placed near an
ideal plane mirror.\cite{Hinds}

\begin{figure}
\includegraphics{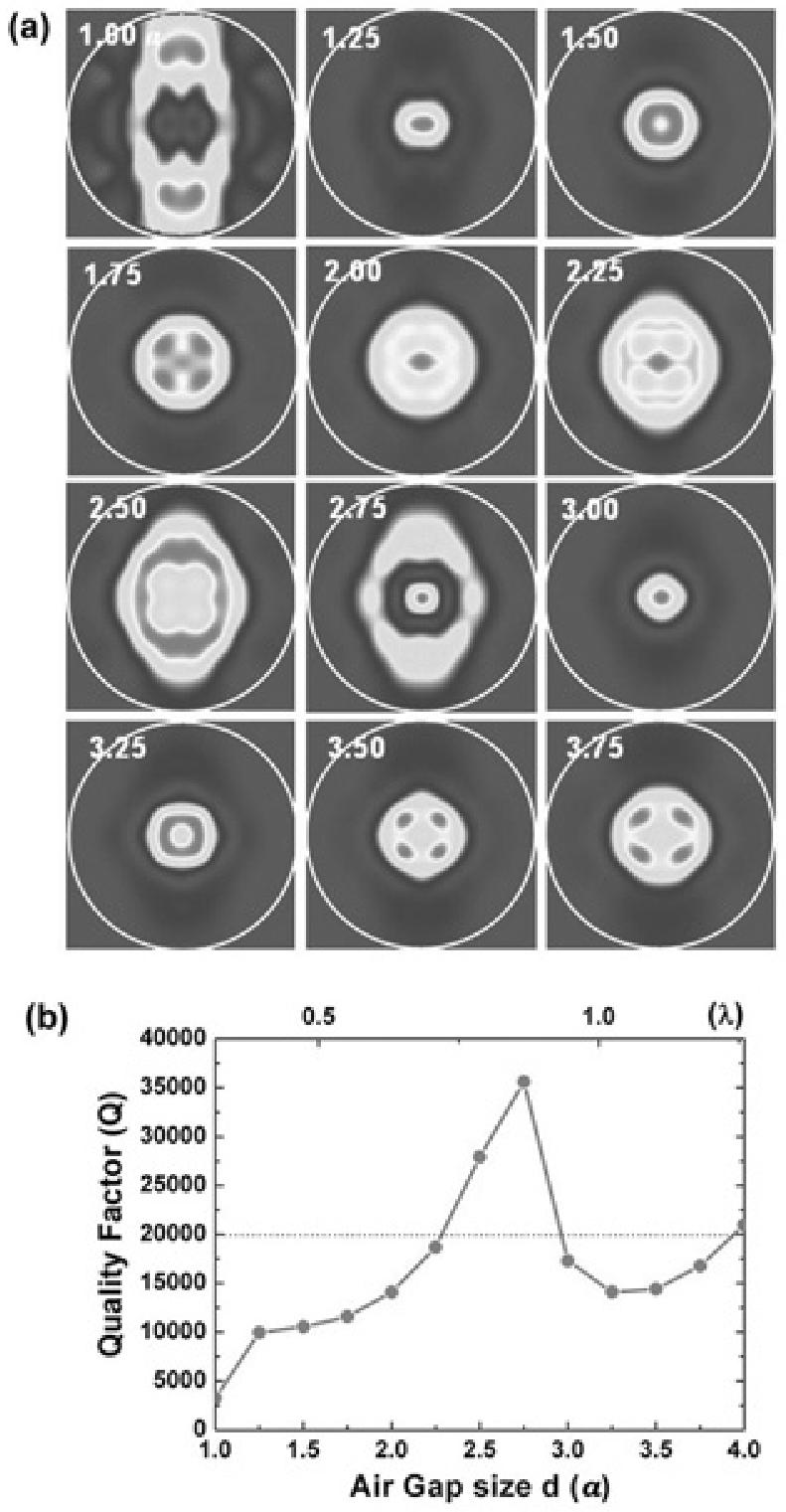}
\caption{\label{fig:fig12} Radiation characteristics of the
deformed hexapole mode in the presence of a Bragg mirror. (a)
Calculated far-field patterns for various values of the air-gap
thickness. Here, each intensity value has been normalized to the
maximum value of its own data. (b) Calculated quality factors as a
function of the air-gap size. A horizontal dotted line represents
the original quality factor in the absence of the reflector.}
\end{figure}

The case of the deformed hexapole mode is worth special attention.
Fig.~\ref{fig:fig11}(a) shows the far-field radiation patterns
with a perturbation parameter $p=0.04 a$ (see
Fig.~\ref{fig:fig7}). The same bottom substrate of index 3.4 is
used. Here again, one can see that the original far-field patterns
are modified by concentric ring-shape modulations. Comparing the
patterns of $1.25 a$ and $3.00 a$, one can witness that the
similar vertical enhancement is restored after the air-gap is
increased by additional thickness of $\sim 1.75 a$. This air-gap
thickness corresponds to, approximately, a half-wavelength ($\sim
1.75 a \times 0.3192 \lambda /a \sim 0.56 \lambda$). At this
point, the vertical enhancement condition seems to be summarized
as $2d \sim m\lambda$ , where $d$ is the air-gap size and $m$
integral multiple. Moreover, the air-gap size of $3.00 a$ where
the good directionality is obtained is approximately equal to one
wavelength ($\sim 0.96 \lambda$). If one wants to find the
rigorous conditions of optimal vertical enhancement, the analyses
need to include the effects of the optical thickness of the PhC
slab and the phase change upon reflection. In the next section, we
shall develop a simple, plane-wave based interference model to
explain the vertical enhancement conditions. From the $Q$ factors,
one can clearly see, again, the modification of the intrinsic
radiation lifetime, where the oscillation period around the dotted
line is $\sim \lambda/2$ [See Fig.~\ref{fig:fig11}(b)]. Note that
there seems to be no definite relation between the $Q$ factor
optimization and the vertical enhancement condition; The optical
loss of the cavity ($\sim 1/Q$) is a result of the sum of the
emitted power ($\int (dP/d\Omega) d\Omega $). A simple guess of
the $Q$ by the emission pattern itself can be misleading.

In principle, one can completely block the downward propagating
energy by using a highly-reflective Bragg mirror. For example, a
stack of 25 AlAs/GaAs pairs give a reflectivity in excess of
0.998.\cite{Coldren} In the subsequent analyses, we will use a
Bragg mirror made of a stack of 20 AlAs/GaAs pairs whose
reflectivity is $\sim 99\%$. Remember that the FDTD grid
resolution for the $z$ direction [$(1/20) a$] is, in fact, not
sufficiently fine to fully describe the each Bragg layer (around
10 grids are used to describe one pair of
AlAs/GaAs).\cite{Taflove} As a result, the reflectivity would be
slightly smaller than the actual value.

Far-field patterns in presence of a Bragg mirror are collectively
shown in Fig.~\ref{fig:fig12}(a). In comparison with
Fig.~\ref{fig:fig11}(a), the concentric ring-shape modulation is
more pronounced due to the high reflectivity of the bottom
structure. In the $Q$ factor plot shown in
Fig.~\ref{fig:fig12}(b), the modification of the radiation
lifetime is also prominent. The maximum $Q$ factor is increased
from $\sim$25 000 to $\sim$35 000, which represents an increase of
the radiation lifetime by $75\%$ in comparison with that without a
reflector. Note that, except for such increased interference
effects, all other characteristic features in the far-field
patterns still remain almost unchanged. In short, we confirmed
that both the radiation lifetime and the radiation pattern can be
controlled by simply adjusting the distance between the PhC slab
and the reflector. As shown in the far-field pattern of $3.00 a$,
one can achieve a sharp directional emission in which most of the
emitted power ($\sim 80\%$) is concentrated within $\pm
30^{\circ}$.

\subsection{\label{sec:sec4p3}Design rule for directional emission}

In the preceding subsection, we have shown that, by using the 3-D
FDTD method and the near- to far-field transformation algorithm,
the radiation patterns of the PhC defect modes can be altered by
the bottom reflector. Especially the choice of $\sim 1 \lambda$
air-gap seems to give the best directional emission. In this
subsection, we will try to find conditions for the directionality,
by developing a simple analytic model based on plane-wave
interferences.

\begin{figure*}
\includegraphics{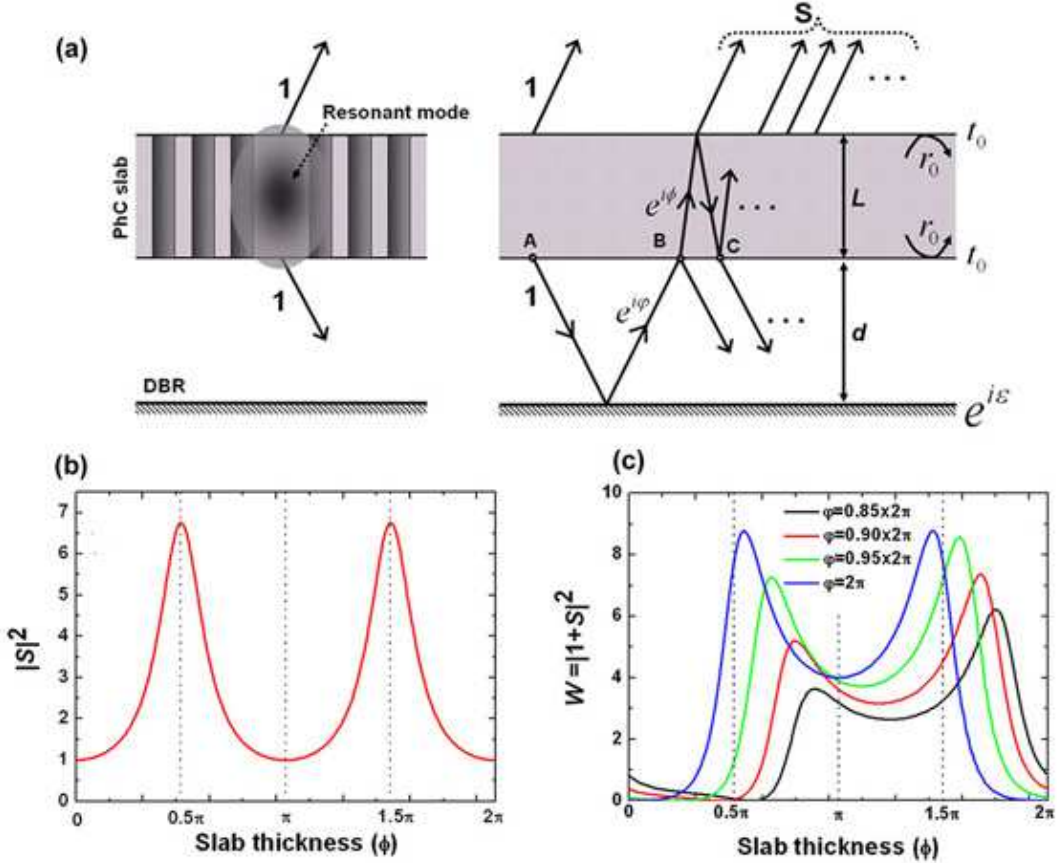}
\caption{\label{fig:fig13} Vertical emission in the presence of
the perfectly conducting mirror (PCM). (a) Simple vertical
interference model. The emission from the cavity is divided by the
upward and the downward propagating waves. (b) Calculated vertical
emission intensity $|S|^2$ ($\theta = 0$) from the initially
downward propagating wave. Here, $\epsilon = \pi$ and $\varphi = 2
\pi$ are assumed. (c) Calculated vertical emission enhancement
factor $W=|1+S|^2$ ($\theta = 0$) as a function of the slab
thickness for various values of the air-gap thickness ranging from
$1.00 \lambda$ to $0.85 \lambda$. }
\end{figure*}

We assume that the PhC slab can be approximated as a uniform
dielectric slab having an effective refractive index\cite{Fan02}
and the PhC resonant mode can be treated as a point dipole source
embedded in the middle of the slab and the radiations are
described by scalar plane-waves. We also assume that the bottom
reflector is positioned sufficiently far ($> \lambda /2$) from the
PhC slab. The reflector does not perturb too much the original
characteristics of the resonant mode. Thus, the initially upward
and downward propagating waves (See Fig.~\ref{fig:fig13}(a)) can
be assumed to be symmetric, 1 and 1. The downward propagating
component will eventually be redirected upward to make an
interference with the upward propagating component. Although these
assumptions seem to be very simple, similar approaches based on
the plane-wave interferences have already been used for the
calculation of radiation patterns of atomic dipoles in a planar
microcavity.\cite{Benisty98,Dowling91} As indicated by
Dowling\cite{Dowling91}, the mode structure of the electromagnetic
field is a classical phenomenon.

In the simplest case when multiple reflections between the PhC
slab and the bottom reflector can be neglected \footnote{When an
optical thickness of a PhC slab is `slab resonance condition', the
PhC slab becomes transparent (zero reflectance).}, a constructive
inteference condition at the vertical becomes

\begin{equation}
e^{2 i \varphi} e^{i\epsilon} e^{i \phi} =1 \label{eq:4p1}
\end{equation}
where $2 \varphi$ is a round-trip phase in the air-gap. $\epsilon$
and $\phi$ are the phase changes at the bottom reflector
\footnote{When we use a Bragg mirror as a bottom reflector, a
reflection phase is $\pi$ at the Bragg frequency. This phase shift
is the same with the case of a perfect conducting mirror.} and
through the PhC slab, respectively (See Fig.~\ref{fig:fig13}(a)).
When the radiation into an angle $\theta$ is considered,

\begin{eqnarray}
\varphi &=& (2\pi/\lambda)d \cos\theta \label{eq:4p2} \\
\phi &=& (2\pi n_{eff}/\lambda) L \sqrt{1-
(1/n_{eff}^2)\sin^2\theta } \label{eq:4p3}
\end{eqnarray}
where $d$ and $L$ are the air-gap thickness and the slab
thickness, respectively. Considering the typical thickness of the
PhC slab of $\approx \lambda /(2n)$, one may assume $\phi \approx
\pi$.\cite{Johnson99} Then, Eq.~(\ref{eq:4p1}) becomes

\begin{equation}
\exp(-i 4 \pi d / \lambda) \approx 1 \label{eq:4p4}
\end{equation}

This condition is consistent with the previous observation ($2d
\sim m\lambda $) which has been deduced from the far-field
patterns. However, in general, the PhC slab thickness does not
satisfy this `slab resonance condition ($L= \lambda/(2n)$)' and
the residual reflection can not be completely cancelled. Then,
what should be changed in the vertical enhancement conditions
[Eq.~(\ref{eq:4p1})] when multiple reflections become significant?

Fig.~\ref{fig:fig13}(a) describes the situation where the emission
from the cavity is divided by the upward and the downward
propagating waves. Here, we focus only on the vertical
interference conditions ($\theta =0$). $r_0$ and $t_0$ are
coefficients of amplitude reflection and transmission for a single
dielectric interface, respectively. Let $S$ be the sum of all the
waves detected upward which initially propagate downward. Then, an
interference between the originally upward propagating wave and
the $S$ will be described by $1+S$, and the resultant intensity
will be proportional to $|1+S|^2$. To sum all the waves which
undergo infinitely many multiple reflections in the slab and in
the air-gap, we introduce a simple method which take advantage of
the similarity of such multiple reflections and transmission.

We trace several reflections and transmissions of the wave which
start out at point $A$ in the initial stage. At point $B$, the
wave is divided into two. Here, the downward propagating component
will undergo the same multiple reflections as the wave at $A$. Let
$Q$ be the sum of all the waves detected upward which propagate
upward at $B$, then we get the following equation for the $S$.

\begin{equation}
S=e^{2 i \varphi} e^{i \varepsilon} \left( t_0 Q + (-r_0) S
\right) \label{eq:4p5}
\end{equation}
Now, let us proceed the upward propagating wave at $B$ more. The
wave once reflected at the top of the slab will reach at point
$C$, there, two divided waves have resemblance to the waves which
contribute to the $Q$ and the $S$, respectively. Describing these
processes,

\begin{equation}
Q=e^{i \phi} \left( t_0 + r_0 e^{i \phi} (r_0 Q + t_0 S) \right)
\label{eq:4p6}
\end{equation}
We get the $S$ by equating Eq.~(\ref{eq:4p5}) and
Eq.~(\ref{eq:4p6}).

\begin{equation}
S = \frac{t_0^2 e^{i \phi}}{ (1-r_0^2 e^{2 i \phi})(r_0 +
e^{-2i\varphi}e^{-i\varepsilon})-r_0 t_0^2 e^{2 i \phi} }
\label{eq:4p7}
\end{equation}

Now, let us examine two representative cases of the PhC slab;
`slab resonance' and `slab antiresonance'.
\\[10pt]
(1) Slab Resonance

This is a case when $\phi = \pi$, i.e., $e^{i \phi} = -1$. From
the well known result for the transmittance in a uniform
dielectric slab,\cite{Coldren} we know that the slab becomes
transparent and multiple reflections in the air-gap do not occur.
Thus, we can easily guess that this case restores the simplest
two-beam interference result dealt in Eq.~(\ref{eq:4p1}). By
inserting $e^{i \phi} = -1$ into the Eq.~(\ref{eq:4p7}), we get

\begin{eqnarray}
S = \frac{t_0^2 (-1)}{
(1-r_0^2)(r_0+e^{-2i\varphi}e^{-i\varepsilon})-r_0t_0^2} \nonumber
\\ = \frac{-1}{e^{-2i\varphi}e^{-i\varepsilon}} =
-e^{-2i\varphi}e^{-i\varepsilon} \label{eq:4p8}
\end{eqnarray}

Resultant radiation intensity becomes

\begin{figure}
\includegraphics{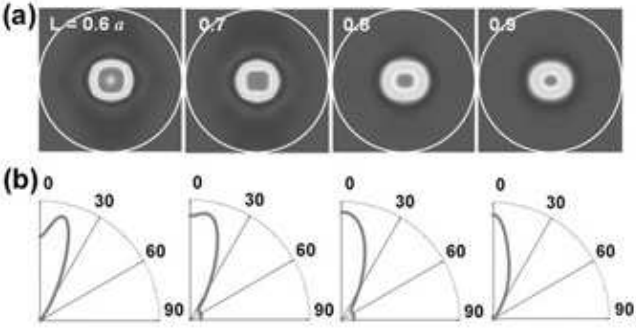}
\caption{\label{fig:fig14} Calculated far-field radiation of the
deformed hexapole mode in the `$\lambda$' air-gap condition. (a)
Far-field patterns for various values of the slab thickness
ranging from $0.6 a$ to $0.9 a$, which approximately correspond to
from $\sim \pi$ to $\sim 1.2 \pi$. (b) Polar plot of the far-field
pattern. Note that the far-field pattern becomes more directional
as the slab thickness increases. From the far-field pattern of
$L=0.9 a$, it is found that more than $80\%$ of total emitted
power is directed within a small divergence angle of $\pm
30^{\circ}$.}
\end{figure}

\begin{equation}
W = |1+S|^2=|1-e^{-2i\varphi}e^{-i\varepsilon}|^2 \label{eq:4p9}
\end{equation}

This is the famous two-beam interference formula, if the phase
change at the bottom mirror is $\epsilon = \pi$, then the
condition for a constructive interference becomes

\begin{equation}
2 \varphi = 2 m \pi \label{eq:4p10}
\end{equation}

The maximum radiation intensity is $W_{max} = 4$ with the above
condition.
\\[10pt]
(2) Slab Antiresonance

This is a case  when $\phi = 1.5 \pi$, i.e., $e^{i \phi} = -i$.
The reflectance from the dielectric slab is maximized with this
slab thickness. Inserting $e^{i \phi} = -i$ into
Eq.~(\ref{eq:4p7}),

\begin{equation}
S = \frac{-i t_0^2}{ (1+r_0^2)(r_0 + e^{-2i\varphi}
e^{-i\varepsilon} ) + r_0 t_0^2 } \label{eq:4p11}
\end{equation}

Here, we let $\epsilon$ again be $\pi$, then one can easily prove
that the condition of the air-gap to maximize $|S|$ is $e^{-2i
\varphi} =1$ (air-gap resonance).

\begin{eqnarray}
S_{max} = \frac{-i t_0^2}{ (1+r_0^2)(r_0 -1) + r_0 t_0^2 }
\nonumber \\
= i \frac{ (1-r_0^2)}{(1-r_0)^2} = i
\frac{(1+r_0)}{(1-r_0)} \label{eq:4p12}
\end{eqnarray}

Assuming the effective refractive index of the slab ($n_{eff}$) by
2.6, then $r_0=(n_{eff}-1)/(n_{eff}+1)\approx 0.44$. The vertical
radiation intensity becomes

\begin{equation}
W = |1+S|^2 = 1^2 + \left( \frac{ 1+r_0}{1-r_0} \right)^2 \approx
7.9 \label{eq:4p13}
\end{equation}
which can be larger than that of the slab resonance. Note that the
condition for maximizing $W$ is slightly different from the slab
antiresonance, because of a phase difference between two complex
numbers ($S_{max}$ is a pure imaginary while `$1$' is a real).
\\[10pt]

In Fig.~\ref{fig:fig13}(c), we plot $W$ as a function of the slab
thickness ($\phi$) for various air-gap sizes ($\varphi$). A blue
line represents the result when the air-gap size exactly equals
the emission wavelength ($d=1.00 \lambda$). Other lines represent
the air-gaps as indicated. Here, it is found that 1) the overall
emission intensity decreases as the air-gap is detuned from the
`$\lambda$' condition and 2) when the air-gap size is
`$\lambda$'(on-resonance), there exists a relatively wide concave
plateau in the neighborhood of the `slab resonance' ($\phi=\pi$).
In this case, the vertical emission is enhanced by a factor larger
than 4, over a wide range of slab thicknesses. Although the better
vertical enhancement can be obtained near the `slab antiresonance'
($\phi=1.5 \pi$), the thickness of the PhC slab is better to be
chosen near the `resonance' condition, because the thicker slab
tends to support unwanted higher-order guided modes.\cite{HYRyu00}
Thus, thanks to the multiple reflection effect, the vertical
enhancement condition now becomes rather simple. Above analyses
suggest that the focus should be placed more on the air-gap size
that has be exactly matched with the emission wavelength $\lambda$
than on the PhC slab thickness that is already designed near the
antiresonance condition.

To confirm the above argument, we have calculated far-field
patterns of the deformed hexapole mode using the `$\lambda$'
air-gap. We have increased the slab thickness from $0.6 a$ to $0.9
a$, which correspond to approximately the phase thickness from
$\sim \pi$ to $\sim 1.2 \pi$ \footnote{Optical thicknesses of the
PhC slab are estimated from the transmittance spectrum which has
been obtained by using the 3-D FDTD method with periodic boundary
conditions. For references, see Fan {\it et al.} \cite{Fan02}}.
Since resonant mode frequencies are changed by the slab thickness,
the air-gap sizes have been adjusted to fit the `$\lambda$'
condition. As shown in Fig.~\ref{fig:fig14}, good directional
patterns are obtained. It is interesting to note that the
far-field pattern becomes more directional as the slab thickness
increases. Therefore, as expected in the previous argument, the
slab slightly thicker than that satisfying the resonance condition
is preferable for the enhanced vertical beaming. From the
far-field pattern of $L=0.9 a$, it is found that more than $80\%$
of the emitted power is contained within an angle of $\pm
30^{\circ}$.

\begin{figure}
\includegraphics{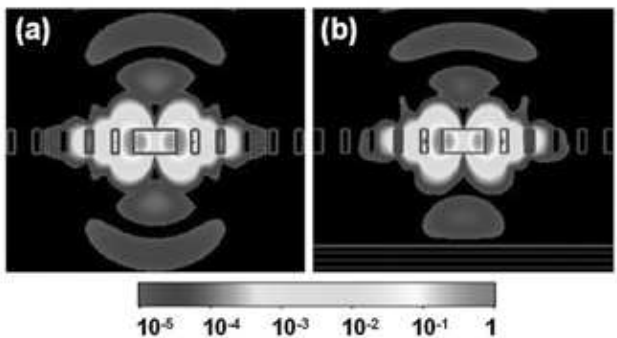}
\caption{\label{fig:fig15} Electric-field intensity distributions
of the deformed hexapole mode detected in the $x$-$z$ plane (a) in
the absence of the reflector (b) in the presence of the Bragg
mirror.}
\end{figure}

Figure~\ref{fig:fig15} shows the electric-field intensity
distributions detected in the $x$-$z$ plane. For comparison, the
case of a symmetric PhC slab cavity without a bottom reflector is
also shown. In Fig.~\ref{fig:fig15}(b), one can clearly see that
the downward propagating waves are effectively redirected by the
Bragg mirror and the unidirectional emission is achieved. In other
words, we have shown that inherent radiation patterns of the PhC
slab resonant mode can be controlled either by modifying the
defect structures, or by placing a reflector near the cavity. By
simply adjusting the air-gap size to be $\lambda$, one can obtain
fairly good directional patterns. The directionality improves
continuously until the slab thickness becomes near the slab
antiresonance. In a practical point of view, the air-gap size is a
fixed quantity which cannot be changed after the wafer growth
\footnote{Recently, LEOM group in France has demonstrated
PC-MOEMS, where one can control the air-gap size by using
electro-static force. See, for instance, J. L. Leclercq, B.
BenBakir, H. Hattori, X. Letartre, P. Regreny, P. Rojo-Romeo, C.
Seassal, P. Viktorovitch, Proc. SPIE, 5450, 300 (2004).}. Thus,
the most important design rule for the directional emission is to
choose the air-gap size (thickness of a sacrificial layer) to be a
target wavelength `$\lambda$'. Then, the remaining critical issue
will be that of the lithographic tuning that matches the resonant
wavelength of the mode of interest with the emission wavelength of
quantum well or quantum dots.
\\

\section{\label{sec:sec5}Summary}

We have shown that the radiation pattern and the radiation
lifetime of the 2-D PhC slab resonant mode can be controlled by
placing a mirror near the cavity. To calculate the far-field
pattern, we have developed a simple and efficient simulation tool
based on the FDTD method and the near- to far-field
transformation. We have discussed two distinct physical mechanisms
responsible for the modification of the far-field radiation
pattern. Either by tuning the defect structure or by placing a
reflector near the cavity, we have shown that the unidirectional
emitter whose energy is concentrated within a small divergence
angle ($\pm 30^{\circ}$) can be achieved. Simple plane-wave based
interference model has revealed that the vertical enhancement
condition is satisfied when the air-gap size is equal to the
emission wavelength. To further enhance the directionality, the
slightly thicker PhC slab design than the `slab resonance'
condition would be recommended. The proposed structure is
practical since the critical design parameter can be precisely
controlled during the epitaxial growth.

\begin{acknowledgments}
One of the authors, S. H. Kim, would like to thank S. H. Kwon, M.
K. Seo, J. K. Yang for their valuable assistance and discussions.
S. H. Kim would also like to thank J. M. G\'{e}rard in CEA in
France for his valuable discussions. This research was supported
by a Grant from the Ministry of Science and Technology (MOST) of
Korea.
\end{acknowledgments}



\end{document}